\documentclass[fleqn,twoside]{article}
\usepackage{espcrc2}

\newcommand{\AmS}{{\protect\the\textfont2
  A\kern-.1667em\lower.5ex\hbox{M}\kern-.125emS}}

\title{Remarks on astrophysical origin of the knee in cosmic ray spectrum}%

\author{Yuri V. Stenkin
\address{Institute for Nuclear Research of Russian Academy of Sciences, \\
        60th October anniv. prosp., 7a, Moscow 117312, RUSSIA,\\
         e-mail: stenkin@sci.lebedev.ru}}%

\begin{document}

\begin{abstract}
All astrophysical explanations of the knee in cosmic ray spectrum
accept the hypothesis of its existence {\it a priori} without any
doubt. But, there exist experimental evidences against this
hypothesis. Experimental data on the knee in various secondary
cosmic ray components obtained by KASCADE array as well as by many
other arrays reveal the existence of the knee
($\Delta\gamma\approx0.4\div0.5$) in electromagnetic component (e
and $\gamma$ including Cherenkov light) and the absence of it or
a very small knee ($\Delta\gamma\approx0.1\div0.2$) in muonic and
hadronic components. But, one could expect the inverse
relationship in case of the astrophysical knee origin. A brief
review of experimental data concerning this problem is given.
\vspace{1pc}
\end{abstract}

\maketitle

\section{Introduction}

The astrophysical origin of the so-called "knee" in primary
cosmic ray spectrum in PeV region is very popular and is widely
exploited in practice for a long time of about 50 years. This
hypothesis seems to have begun a new life in the recent years
after discovering some new Supernova remnants. But, experimental
data accumulated up to now in cosmic ray physics seed some doubt
on the "knee" existence. One could argue that there exists no
theory able to explain pure power law spectrum in a very wide
range from $10^{10}$ to $10^{20}$ eV. But, such a theory does
exist \cite{trub}. It was published many years ago and proposed a
universal mechanism of cosmic ray acceleration in the cosmic
plasma pinches. This theory can explain not only power law
spectrum but even the value of  its power law index $\gamma$. This
mechanism can generate cosmic rays up to the highest energy with
the integral exponent $\gamma=\sqrt{3}\approx1.73$. Another
outgoing feature of this mechanism is as follows: the index does
not depend on  a particle mass or charge. It is really a
universal value.

\section{Brief review of the experimental data}

Let us look at the problem from the experimental point of view
\cite{sten1}. Direct measurements do not show any significant
change in the spectrum slope up to energy 1 PeV. The "knee" in
the size spectrum is only seen in secondary cosmic ray components
measured by the indirect method of Extensive Air Shower (EAS). If
primary spectrum index is $\gamma$ and a secondary x-component
depends on primary energy $E_{0}$ as $N_{x}(E_{0})\sim
E_{0}^{\alpha}$ then the EAS-size distribution on $N_{x}$ is
$P(N_{x})\sim N_x^{-\beta}$, where $\beta=\gamma/\alpha$. If the
"knee" in primary spectrum does exist (let it be
$\Delta\gamma=0.5$) one can predict with rather high accuracy a
relationship between the "knees" in all detectable secondary
components: electromagnetic, muonic and hadronic. Index $\alpha$
commonly used for electrons is equal to $\alpha_{e}=1.15-1.25$ (so
$\Delta\beta_{e}=\Delta\gamma/\alpha_{e}\approx0.42$) and that for
muons and hadrons is equal to $\alpha_{\mu,h}=0.8-0.95$.

\begin{table*}[htb]
\caption{Measurements, predictions or expectations for the knee value}
\newcommand{\m}{\hphantom{$-$}}
\newcommand{\cc}[1]{\multicolumn{1}{c}{#1}}
\renewcommand{\tabcolsep}{0.35pc} 
\renewcommand{\arraystretch}{1.1} 
\begin{tabular}{@{}lllll}
\hline
$\Delta\beta_e$ & \cc{$\Delta\beta_\mu$} & \cc{$\Delta\beta_h$} & %
\cc{$\Delta\gamma$} & \cc{Reference} \\
\hline
$0.42$   & \m0.57  & \m0.57   & \m$\equiv0.5$ & \m(in a case of knee) \\
$\approx0.47$ (\v{C}) &\m-  & \m- & \m-  & \m TUNKA$\cite{bud}$  \\
0.57 (e); 0.52 (\v{C}) &\m- &\m- & \m$0.5\pm0.6$ &\m HEGRA$\cite{arq}$\\
$\approx0.40$  &\m-  & \m- &\m$0.40\pm0.06$ &\m MAKET-ANI$\cite{chi}$\\
$- $    & \m$\approx0.2$ for all & \m- & \m$0.2\pm0.02$ for all & %
\m KASCADE$\cite{ant}$ \\
$- $    & \m0 for "e-poor"  & \m- & \m0 for "e-poor" & \m KASCADE$%
\cite{ant}$ \\
$- $    & \m- & \m$0.11\pm0.02$ for $\Sigma$$E_{h}$ & \m- & \m KASCADE$%
\cite{hor}$ \\
$- $  & \m- & \m$0.2\pm0.1$ for $N_{h}$ & \m- & \m KASCADE$\cite{hor}$\\
$0.35\pm0.1$ (vertical)& \m0 & \m- & \m $0.40\pm0.12$ & %
\m AKENO$\cite{nag}$\\
$0.42\pm0.03$ & \m- &\m- &\m$\approx0.4\div0.5$ &\m MSU EAS$\cite{fom}$\\
$- $  & \m$\sim0$ & \m- & \m- & \m MSU EAS$\cite{fk}$ \\
$0.43\pm0.09$  & \m$0.4\pm0.1$ & \m- & \m$0.4\pm0.1$ & \m EAS-TOP$%
\cite{agli}$ \\
$- $    & \m$\approx0$  & \m-  & \m-   & \m GAMMA$\cite{chi2}$ \\
$- $    & \m$\sim0$ & \m- & \m$0$ & \m underground $\mu \cite{sten2}$ \\
$- $    & \m$\sim0$  & \m-  & \m-  & \m DELPHI$\cite{trav}$  \\
$- $    & \m$\sim0$  & \m-  & \m-  & \m ALEPH$\cite{ava}$  \\
$\approx0.5$ &\m$\sim0$ &\m$\sim0$ &\m$0$ &\m prediction$\cite{sten1}$\\
\hline
\end{tabular}
\end{table*}

Therefore, the visible "knee" ($\Delta\beta_{\mu,h}\approx
0.5/0.87\approx0.57$) in muonic and hadronic components is
expected to be bigger than in electrons (in EAS size spectrum)
and even bigger than in primary spectrum, because
$\alpha_{\mu,h}<1$ while $\alpha_{e}>1$. It is a very strict
limitation because $\alpha_{\mu,h}$ and $\alpha_{e}$ were
measured in many experiments and were calculated by many authors
using different Monte-Carlo programs. In this paper I would like
to emphasize that experimental data on a visible "knee" in
different secondary components contradict  this limitation. In
other words, the values of the "knee" in primary cosmic ray
spectrum reconstructed from different secondary components are
not equal and contradict each other as is seen in Table 1. In all
experiments shown above
$\Delta\beta_{e}>\Delta\beta_{\mu}\approx\Delta\beta_{h}$. The
most detailed and careful measurements in the knee region have
been performed by KASCADE experiment specially designed to solve
this problem. The "knees" claimed by the authors %
in muonic and hadronic components are much less than expected ones
(even equal to 0 for 'e-poor' EAS's) and are very smooth and thus
could be regarded from my point of view as a smooth change of the
power index in a very wide range due to methodical or other
reasons to be discussed below. Therefore, their values could be
regarded as the upper limits for the "knee". If so, the KASCADE
data are consistent with the "zero knee" hypothesis made on the
basis of all available data analysis \cite{sten2} of the deep
underground muon experiments (Baksan, Frejus, NUSEX, MACRO,
SOUDAN2). In fact, only 2 experimental groups have confirmed the
existence of the knee in the EAS muonic component comparable with
that for the electron component: MSU EAS and EAS-TOP. But their
data need special comments. MSU EAS muon data obtained with muon
detectors do not reveal a knee. What is usually shown by the
authors is a muon number spectrum \emph{recalculated from the
electron number spectrum}. Therefore, it must have the knee close
to that in EAS size spectrum. As for EAS-TOP data, their muon
number spectra shown in $\cite{agli}$ (and in other publications
as well) could be successfully fitted by a pure power law by
taking into account statistical errors and the points spread.
Moreover, the results of these two experiments contradict each
other because EAS-TOP claimed the knee in muonic component
($E_{\mu}>1$ GeV) at $N_{\mu}=10^{4.9}$ while MSU EAS group
($E_{\mu}>10 GeV$) do not see any knee above $N_{\mu}=10^{4}$
(the difference in muon threshold energy gives a factor of 5-6
while the difference in figures is higher and the difference in
altitude of $\sim200 g/cm^{2}$ is negligible for muons). The
result of AKENO combined in \cite{nag} with SUGAR data \cite{sug}
gave a rather strict limitation for muon size spectrum as
follows: " ...there is no change of slope in muon size spectrum
between $10^{5}$ and $10^{9}$ ". Another strict limitation
starting from a very low muon size has been given by GAMMA data
\cite{chi2} for muons with energy above 5 GeV: "The spectrum
above $N_{\mu}>7000$ is well described by the power law
dependence $F(>N_{\mu})\sim N_{\mu}^{-2.18}$ ".

An absolute value of the power law index $(\beta_{\mu})$ has to
be discussed separately. As it was demonstrated in \cite{sch} the
indices of primary spectrum derived from measurements by different
experimental groups are in disagreement. In our approach
$\gamma=const$ and everything is explained by the index $\alpha$
behaviour, which is not constant in different experiments. This
can be explained at least for the measurements of primary spectrum
through the muonic component. As it has been shown \cite{chu}
muon number spectrum is connected with the primary cosmic ray
spectrum and follows power law if primary spectrum does so. In
case of an infinite area detector, the simplest one corresponding
to the total number of muons in EAS when
$\alpha_{\mu}\approx0.7\div0.8$ (let it be equal to 0.7 for
estimations), the index of muon number distribution in asymptotics
should be equal to
$\beta_{\mu}=\gamma/\alpha_{\mu}\approx1.7/0.7\approx2.4$. Low
multiplicities $(m<5)$ are produced by low energy primaries and
can be described by Poissonian or binomial distributions
characteristic for the discrete values. Such a spectrum was
observed in many underground experiments (see for example
\cite{sten2}) using large area detectors for recording of muons
with a high energy threshold without any EAS trigger (case (a)).
A similar result is expected for small underground detectors but
only the ratio of single muon flux (m=1) to that for muons in
groups $(m>1)$ is expected to be much higher. Another interesting
feature of such experiments (without EAS core location) is the
so-called {\it core attraction effect} \cite{bak}: the higher
multiplicity in the detector is - the sharper recorded EAS core
lateral distribution centered in the detector is. This effect
results in a measurement of not the total number of muons
integrated over all distances up to infinity but of only the
central muon density having a pole in the axis where the exponent
is equal to $\alpha_{\mu}\approx1$. Therefore, the measured muon
number spectrum at very high multiplicity (where the pole does
exist) should be flatter \cite{sten3} having the index
$\beta_{\mu}=\gamma/\alpha_{\mu}\approx1.7/1.0=1.7$. The latter
could probably explain the problem with the highest multiplicity
events observed in the CERN detectors ALEPH and DELPHI. In real
underground experiments due to different experimental and trigger
conditions and due to different statistics, the measured integral
index should lie between these two extreme values
$(\beta\approx2.4\div1.7)$.

The opposite case (b) occurs when the muon detector is working
under the EAS array trigger. If the data are corrected for the
trigger bias then the EAS muon size spectrum will follow the power
law with $\beta\approx2.4$. If no corrections are applied then
the measured muon multiplicity spectrum is very flat at low
multiplicity ($\beta\approx0.7$) \cite{chu,alex,bak} according to
the Poissonian distribution with the high mathematical
expectation $(m_{0}>1)$ which depends on the EAS size threshold
and on the muon detector total area. Only at a high enough number
of muons, the spectrum becomes as steep as in case (a). This
effect could imitate the expected knee in primary spectrum. If
muon density is measured at fixed core distances (as in KASCADE
and many other EAS experiments) then the core attraction effect
does not work and no flattening occurs at high multiplicity in
such experiments. Here could be probably found the difference in
experimental data at the highest muon density measurements
between EAS arrays and underground detectors.

 Unfortunately there exist very few data on the
hadronic EAS size spectrum measurements free of trigger
conditions bias. I found only one experiment (\cite{hor})
providing such data.

We can conclude that none of the experiments has shown undoubtedly
the knee existence in muonic or hadronic component in the needed
place and of the needed size as could be expected in case of real
knee existence in primary spectrum.\footnote{concerning the
KASCADE muonic data this has been first mentioned in \cite{ili}.}

\section{Summary}

Summarizing the experimental data shown above we can conclude that
the question "Does the knee exist or not?" is still open.
Therefore, the main efforts should be directed now not to
explanations of the "knee" but to the question above. Only when
this question is answered positively an astrophysical explanation
of the "knee" existence in primary cosmic ray spectrum will be
needed. Up to date I incline to a negative answer to this
question, because only in this case we could understand the
experimental data. In the frame of the approach proposed in
\cite{sten1} there exists a natural explanation of the visible
"disagreement" of the experimental data shown in Table 1: the
"knee" does not exist in primary cosmic ray spectrum and should
be observed only in electromagnetic component due to non-power
law behaviour of the $N_{e}(E_{0})$ in a region of $\sim100$
TeV/nucleon when hadron core reaches the sea level and {\it
coreless EAS's} become {\it coreful} or normal ones. Muons and
hadrons are almost insensitive to this critical point of the EAS
development and follow a power law much better.

Looking at Table 1 we can conclude that the measured data are more
consistent with the prediction \cite{sten1} shown in the last
line than  with the expectations made above under a supposition
of the "knee" existence (first line). And finally: if the answer
is negative then we probably have a natural universal theory of
cosmic ray acceleration [1]. To confirm or to reject this theory
we have to measure the spectrum index $\gamma$ with high accuracy
and answer the question: "Does it equal to $\sqrt{3}$ or not?"

The author thanks Dr. A.L.Tsyabuk for fruitful discussions of
muonic data.

This work was supported in part by the RFBR grants N 02-02-17290 %
and N 04-02-17083.

\end{document}